\documentclass[11pt]{emulateapj}

\def \Spitzer{{\emph{Spitzer}}}

\def \um {{$\mu$m}}

\def \msun {{$M_{\odot}$}}

\begin{document}
\slugcomment{12/11/12}

\title{The Contribution of Thermally-Pulsing Asymptotic Giant Branch and Red Super Giant Stars to the Luminosities of the Magellanic Clouds at $1-24$~\um}

\author{J. Melbourne\altaffilmark{1} \& Martha L. Boyer\altaffilmark{2, }\altaffilmark{3}}%, Mark Seibert\altaffilmark{3} }
\altaffiltext{1}{Caltech Optical Observatories, Division of Physics, Mathematics and Astronomy, Mail Stop 301-17, California Institute of Technology, Pasadena, CA 91125, jmel@caltech.edu}
\altaffiltext{2, }{Observational Cosmology Lab, Code 665, NASA Goddard Space
        Flight Center, Greenbelt, MD 20771, USA; martha.l.boyer@nasa.gov}
\altaffiltext{3}{Oak Ridge Associated Universities(ORAU), Oak Ridge, TN 37831, USA}

%\altaffiltext{3}{Observatories of the Carnegie Institution of Washington, 813 Santa Barbara St., Pasadena, CA 91101, }

\begin{abstract}
We present the near- through mid-infrared flux contribution of
thermally-pulsing asymptotic giant branch (TP-AGB) and massive red super giant (RSG) 
stars to the luminosities of the Large and Small
Magellanic Clouds (LMC and SMC, respectively). Combined, the peak contribution
from these cool evolved stars occurs at $\sim3-4$~\micron, where they produce 
32\% of the  SMC light, and 25\% of the LMC flux.
The TP-AGB star contribution also peaks at $\sim3-4$~\micron\ and amounts to 21\%  in both 
galaxies. The  contribution from RSG stars  peaks at
shorter wavelengths, 2.2~\micron, where they provide 11\% of the SMC  flux, and 7\% for the LMC. 
Both TP-AGB and RSG stars are short lived, and thus potentially impose a large stochastic scatter on
the near-IR derived mass-to-light ratios of galaxies at rest-frame 1--4~\micron.  
To minimize their impact on stellar mass estimates, one can use the M/L ratio at
shorter wavelengths (e.g. at 0.8 - 1~\micron). 
At longer wavelengths ($\geq$8~\micron), emission from dust
in the interstellar medium dominates the flux. In the LMC, which shows strong PAH emission at 8~\um, 
TP-AGB and RSG contribute less than 4\% of the 8~\um\ flux.  However, 19\% of the 
SMC 8~\micron\ flux is from evolved stars, nearly
half of which is produced by the rarest, dustiest, carbon-rich TP-AGB stars. Thus, star
formation rates of galaxies, based on an 8~\um\ flux (e.g. observed-frame 24~\micron\ at $z=2$),
may be biased modestly high, especially for galaxies with little PAH emission.

\end{abstract}


\section{Introduction}

%{\bf Note: I edited the title, since it seemed long.  But maybe we can
%  re-add a subtitle back in.  Also, AGB and TP-AGB are not necessarily
%  interchangeable, so I've changed most occurrences of TP-AGB with
 % just AGB.  It's likely that there are plenty of faint AGB stars that
 % don't quite qualify as TP-AGB. But I know that most people in the
 % literature use TP-AGB, so I'm open to calling them all TP-AGB.}

Stellar masses of galaxies are typically estimated from a model
mass-to-light (M/L) ratio \citep[e.g.][]{Bell01}, and a measurement of
the integrated luminosity at optical or near-infrared wavelengths.
Until recently, galaxy M/L ratios were thought to be relatively well
behaved, especially at near-infrared wavelengths where the effects of
dust and massive main sequence stars are minimized.  However, in the
past decade this assumption has been shown to be flawed because of the
influence of short-lived but extremely luminous thermally-pulsing asymptotic giant
branch (TP-AGB) stars \citep[][$0.8 \lesssim M \lesssim 8\,
  M_\odot$]{Maraston06, Zibetti09} and massive red super giant stars 
  \citep[RSG;][]{Dalcanton12a, Melbourne12a}.

Just as the near-IR fluxes of galaxies are often used to estimate
their stellar masses, mid-IR fluxes can be used to estimate their star
formation rates.  For instance, at rest-frame 8 \micron, the flux from
starbursts is typically dominated by polycyclic aromatic hydrocarbon (PAH) emission which scales with
the star formation rate \citep[SFR; e.g.,][]{Diaz-Santos08}.  Likewise the
rest-frame 24 \micron\ flux has been shown to scale with the IR
luminosity of a galaxy and is therefore used as a SFR indicator
\citep{Chary01}.  However, calibrations of these SFR indicators
typically ignore the possibility of stellar contamination in the flux at these wavelengths.
Several works have shown that, under certain assumptions, AGB stars
could contribute significantly to the mid-IR fluxes of galaxies and
therefore affect the SFRs estimated from their mid-infrared 
fluxes \citep[e.g.,][]{Verley09, Kelson10, Chisari12}. However, as will be discussed in this paper,
some of their assumptions may be flawed.

The AGB is a short-lived phase \citep[$\sim1$ Myr, e.g.,][]{Girardi10} near the end of stellar evolution when a star undergoes helium shell burning and swells to a large size, producing a relatively cool photosphere and, potentially, a very infrared- (IR-) luminous star. Because of the cool temperatures, complex molecules and dust can form in the atmosphere \citep[e.g.,][]{Bowen88, Winters00, Winters03, Schirrmacher03, Woitke06, Woitke07, Mattsson08, van-Loon08}, which in turn can drive strong stellar winds \citep[e.g.,][and references therein ]{Sedlmayr95, Elitzur01} that ultimately return as much as half of the stellar mass to the interstellar medium.  An isolated star will start its AGB phase with an oxygen-rich atmosphere (e.g.,O-AGB). However, during the later stages of AGB evolution, a star undergoes a series of thermal-pulses and carbon, dredged up from the interior, pollutes the atmosphere \citep{Iben83a}.  When the C/O ratio exceeds unity the star is classified as a carbon star (e.g., C-AGB), and complex molecules and dust grains form in the escaping winds. These dust grains tend to obscure the star at optical and near-IR wavelengths, but the same dust emits in the mid-IR. Only 20\% of the AGB stars in the LMC/SMC are estimated to be thermally-pulsing (Bruzual, private communication). However, these TP-AGB stars can be very luminous in the infrared, typically exceeding the luminosity of the tip of the red giant branch (TRGB), sometimes by several magnitudes.      

Red supergiants are helium-burning massive stars that form a tight sequence in optical and near-IR CMDs (e.g. Figure \ref{fig:CMD}).  Stars on this sequence are also termed red helium burning stars \citep[RHeB][]{Dohm-Palmer02,  McQuinn11, Dalcanton11}.  At the luminous end, RSGs are truly massive, 9 solar masses or more and thus were born in very recent star formation episodes (e.g. $< 0.1$ Gyr).  At the luminosity of the $K$-band TRGB, the Padova isochrones predict more modest masses, $6-7$ \msun\ \citep{Bertelli08, Marigo08, Bertelli09, Girardi10}, and in fact the RHeB sequence continues to lower luminosities for stars of even smaller mass \citep[e.g.][]{McQuinn11}.    

Combined, TP-AGB and RSG stars have been shown to account for as much
40\% of the integrated 1.6~\micron\ flux, even in local galaxies, where
the red giant branch (RGB) is well developed \citep{Melbourne12a}.  In
the early universe, when the RGB has not had time to form, the TP-AGB and
RSG stars are expected to contribute even larger fractions (as much
as 70\%) of the near-IR light, even though they represent a tiny
fraction of the stellar mass \citep{Maraston06, Melbourne12a}. 
Thus, these stars must be accounted for to accurately
model the infrared M/L ratios of galaxies \citep{Zibetti09, Ilbert10}.

Not only do these stars contribute to the near-IR luminosities of
galaxies, but they also can be significant sources of mid-IR
light \citep[e.g.,][]{Boyer11}.  TP-AGB stars, in
particular, can produce significant amounts of warm dust 
\citep[e.g.,][]{Bowen88, Schirrmacher03}.  
%The primary driver of the high
%mass loss in TP-AGB stars is pulsation-enhanced, dust-driven winds
%where pulsations produce density enhancements and shocks that
%encourage dust formation in the stellar atmosphere
%\citep[e.g.,][]{Bowen88, Winters00, Winters03, Schirrmacher03,
 % Woitke06, Woitke07, Mattsson08, van-Loon08}.  %Radiation pressure on
%these dust grains drives a stellar wind \citep[e.g.,][and references
%  therein ]{Sedlmayr95, Elitzur01} that can ultimately remove more
%than 50\% of the stellar mass. 
These dusty shells both emit in the mid-infrared and obscure the the TP-AGB stars at optical wavelengths \citep[e.g.,][]{Reid91}.

The flux contribution of TP-AGB and RSG stars has now been well
quantified at 0.8 and 1.6~\micron\ in a set of 23 local dwarfs and spirals
\citep{Girardi10, Melbourne12a}. Similarly the 3.6 and 4.5~\um\ flux contribution of the 
most extreme (dusty) TP-AGB stars has been quantified in a set of six nearby 
galaxies \citep{Gerke12}. Likewise the mid-IR ($8 -24$ \um) flux contribution from 
carbon stars has been estimated from stellar population synthesis models of SINGS
galaxies \citep{Chisari12}. 
However, there has not been a study that quantifies the flux contributions of these stars
across the full $1-24$~\um\ wavelength range for a galaxy where complete samples of TP-AGB 
and RSG stars are actually resolved and individually measured.  

In this paper, we use complete samples of TP-AGB and massive RSG
stars in the Magellanic Clouds from the {\it Spitzer} legacy program
``Surveying the Agents of Galaxy Evolution''
\citep[SAGE,][]{Blum06,Meixner06,Gordon11,Boyer11}. We build off of
the results of \citet{Boyer11}, who showed that the TP-AGB and RSG
stars contribute significantly to the near- and mid-IR point-source
fluxes in these galaxies.  We now show that their contribution to the
total (stellar and interstellar) fluxes of the SMC and LMC varies
significantly from 1 to 24~\micron.  We also find significant differences between our 
measured flux contributions and those from several recent studies in the literature. We find a significantly larger 
TP-AGB contribution at 3.6~\um\ than is found in \citet{Gerke12} and \citet{Meidt12a}.  Conversely, we find the $8-24$~\um\ AGB flux contributions estimated in \citet{Kelson10} and \citet{Chisari12} are likely significantly over-estimated.
We discuss the implications of these results for estimates of stellar mass and SFRs of galaxies. 
    
\section{The Data}
This paper uses several types of observations: (1) ground-based,
seeing-limited near-infrared imaging to measure the near-infrared
fluxes of TP-AGB and RSG stars in the Magellanic Clouds, (2)
space-based, low-resolution near-infrared imaging to determine the
total integrated near-infrared luminosities of the Magellanic Clouds, and (3)
space-based, mid-infrared imaging to determine the mid-infrared fluxes
of both the stars and the integrated galaxy light.  In addition to the IR data,
we use ground-based optical data to explore the flux contribution of short-lived 
massive main sequence stars at shorter wavelengths. Details of these
datasets are provided below. All magnitudes are reported in the Vega system.
  
\subsection{Stellar Photometry}
Multi-wavelength stellar photometry of the stars in the Large and Small Magellanic
Clouds (LMC and SMC, respectively) was compiled by the SAGE team.
$J$ and $K_{\rm S}$ (ie., 1.2 and 2.2 \micron)
near-infrared photometry was obtained from the Two Micron All Sky
Survey \citep[2MASS;][]{Skrutskie06}. Mid-infrared photometry at 3.6,
4.5, 5.8, and 8 \micron\ was obtained with the {\it Spitzer Space Telescope} InfraRed
Array Camera \citep[IRAC;][]{Fazio04}. Photometry at 24 \micron\ was
obtained with the Multi-band Imaging Photometer for {\it Spitzer}
\citep[MIPS;][]{Rieke04}. A discussion of the {\it Spitzer}
observations, image reductions, photometry, and source matching for
the SMC is given in \citet{Gordon11}, and the LMC photometry is
discussed in \citet{Meixner06}. Optical, $BVI$, data was compiled by the Magellanic Clouds
Photometric Survey \citep[MCPS;][]{Zaritsky02}.

\begin{figure*}
\centering
\plotone{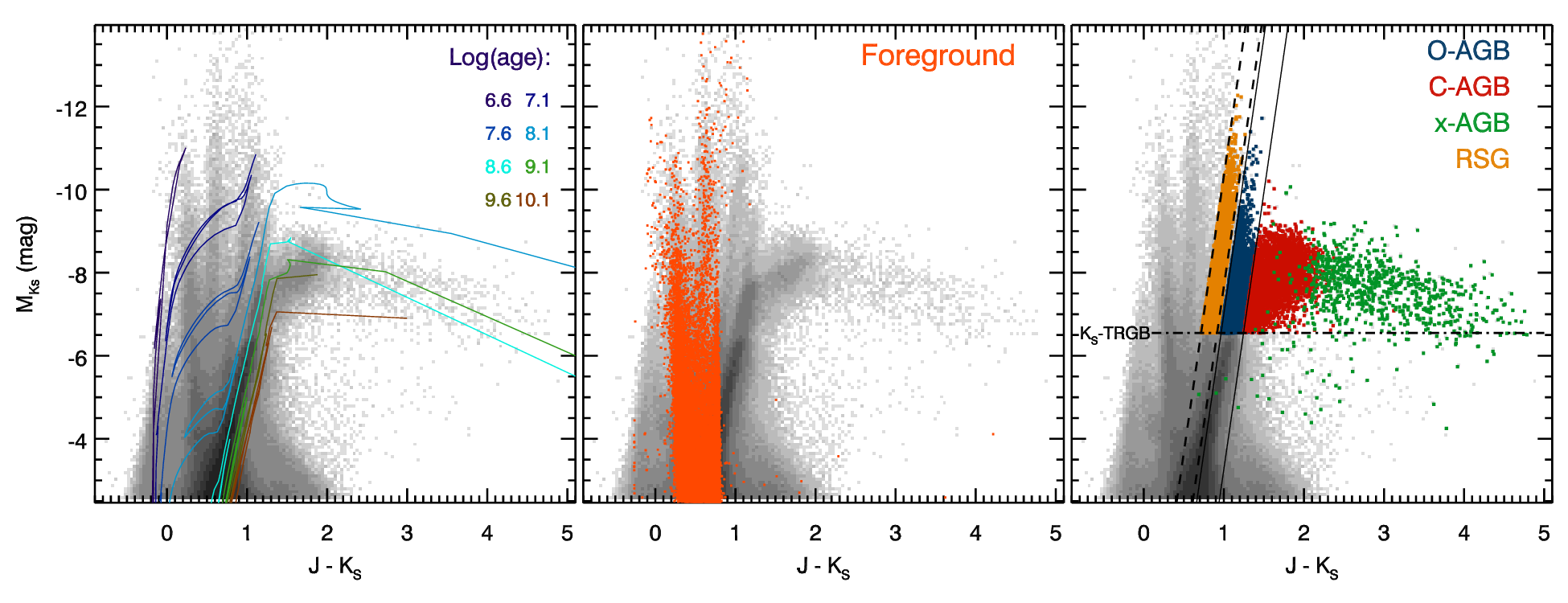}
\caption{\label{fig:CMD} $J-K_{\rm S}$ vs. $K_{\rm S}$ CMDs for the LMC (grey density plots).  Left panel overlays the Padova isochrones \citep{Bertelli08, Marigo08, Bertelli09, Girardi10}.  Center panel overlays a statistical estimate of the Milky Way foreground stars in the direction of the LMC from TRILEGAL \citep{Girardi05}. The right panel shows the \citet{Boyer11} classification of RSG (yellow), O-AGB (blue), and C-AGB (red) stars, with selection regions defined by \citet{Cioni06}.  A luminosity cutoff was made at the TRGB, below which the fluxes of these stars become negligible compared with the RGB. The dust enshrouded x-AGB (green) were selected by \citet{Boyer11} at longer wavelengths; $J-K_{\rm S} > 3.1$ mag or $[3.6]-[8] > 0.8$ mag and $[8] < 10$ mag.   Because of their dusty envelopes, the x-AGB stars can be fainter than the $K_{\rm S}$-band TRGB and yet be significantly brighter than the TRGB at longer wavelengths. 
}
\end{figure*}

Stellar classifications for stars in the SAGE data were presented in \citet{Boyer11}.  Here we briefly summarize the method; for a full description of the stellar classification and sources of possible contamination see \citet{Boyer11}.  Classification was carried out by examining the locations of stars in multi-wavelength color-magnitude space. Less-dust-obscured TP-AGB and RSG stars were identified in near-infrared color magnitude diagrams (CMDs), where they are among the most luminous stars, e.g. above the $K_{\rm S}$-band TRGB.  At these wavelengths, sequences of carbon-rich and oxygen-rich AGB stars (C-AGB and O-AGB stars respectively) separate by $J-K_{\rm S}$ color, with C-AGB stars typically redder than O-AGB stars.  Likewise, RSG stars form a tight sequence slightly blueward of the O-AGB stars.  These sequences are clearly visible in the $J-K_{\rm S}$ CMD of the LMC shown in Figure \ref{fig:CMD} which also shows the color-magnitude boundaries used to identify these different sequences \citep[adopted from][]{Cioni06}.  While these boundaries are useful for identifying different classes of stars, we note that there will be some overlap between the populations. Boyer et al. (2011) provides a detailed discussion of classifications including estimates of the numbers of misclassified stars, which are small. We note that the anomalous O-rich AGB stars reported by \citet{Boyer11} are included with the O-AGB sample here.

The most dust enshrouded TP-AGB stars are significantly extinguished even in the near-IR, and therefore they must be identified at longer wavelengths \citep[e.g.,][]{Reid91}.  \citet{Boyer09} found that short wavelength searches can miss up to 40\% of the TP-AGB stars.  These so-called extreme AGB stars (x-AGB) were identified by their colors at mid-IR wavelengths (3.6 to 8~\um).  Most x-AGB stars are C-rich \citep{van-Loon05}.  Combining these two identification methods allows for the selection of near-complete samples of TP-AGB stars (e.g. these classifications do not miss the most dust enshrouded stars, which can be missed by classifications that rely entirely on shorter wavelengths). 

While these techniques will recover the bulk of the luminous TP-AGB stars, some TP-AGB stars will be missed because of they are at a minimum in their thermal pulsation cycle and are fainter than the TRGB. Models of the TP-AGB stars in the Magellanic Clouds based on the Padova stellar evolution tracks reveal that $<10$\% of them lie below the TRGB due to the thermal pulse cycle \cite[e.g.][]{Marigo07}.  We have included these sub-luminous TP-AGB stars in the uncertainties of the integrated TP-AGB fluxes by assuming they have the luminosity distribution at the top magnitude of the RGB.   We added this uncertainty in quadrature to the other sources of uncertainty, increasing the upper error bars in Table 1 and Figure 2 by $<3$\%.

We also note that there are AGB stars that are not thermally-pulsing.  These are lower-luminosity AGB stars and also lie below the TRGB.  These stars cannot be easily selected in CMD space because they overlap the RGB stars.  However, as these AGB stars are also relatively rare by number compared with the equally luminous RGB, they will not contribute significantly to the total flux of the galaxy. 

The total numbers of SMC and LMC stars within each stellar classification bin are presented in Table \ref{tab:results}.

\subsection{Foreground Contamination}
The CMDs of the LMC and SMC also contain some Milky Way foreground stars with the colors and luminosities of Magellanic Cloud AGB and RSG stars.  We applied a statistical correction for foreground contamination of Milky Way stars at each wavelength.  We used TRILEGAL \citep{Girardi05}
to model the numbers and fluxes of Milky Way  stars in the direction of the Magellanic Clouds in the same filters as our data. The TRILEGAL photometry was then run through the same classification scheme as the RSG and TP-AGB stars. The contamination is largest for the RSG population of the SMC ($\sim25$\%). In the LMC the RSG contamination is negligible because the higher metallicity of the LMC causes all of the RSG and AGB stars to be redder than they are in the SMC, resulting in less overlap with the foreground sources in the near-IR CMD (see Figure \ref{fig:CMD}). Likewise, the AGB populations in both galaxies are effectively foreground contamination free. The estimated numbers of foreground stars within each stellar class are also presented in Table \ref{tab:results}. 

 Note: stars in the Milky Way foreground clusters 47 Tuc and NGC 362, in the direction of the SMC, were removed from the data \citep[see][]{Boyer11}.

\subsection{Integrated Galaxy Photometry}
Measuring the total luminosity of the Magellanic Clouds is challenging
because of their very large areas on the sky.  Ground-based, near-IR
images are dominated by thermal background flux, and thus accurately
discerning sky from unresolved galaxy becomes nearly
impossible. Fortunately these issues are largely eliminated by going
to space, where the IR thermal background is minimal and COBE/DIRBE
\citep{Hauser98} provides a large beam on the sky (nearly a degree).

We use the COBE/DIRBE integrated flux density estimates of
\citet{Israel10} for the Magellanic clouds at 1.25, 2.2, and 3.5~\um. 
The SMC flux is measured within a 12\fdg3$\times$12\fdg3 area, and
the LMC flux covers a 15\fdg3$\times$15\fdg3 area. These measurements were carefully
corrected for foreground (and background) contamination by \citet{Israel10}. 
The Israel et al. flux density estimates for the Clouds are presented here in Table \ref{tab:results}.

Similarly, the global SMC and LMC IR fluxes from the \Spitzer\ data at 3.6,
4.5, 5.8, 8, and 24~\micron\ were made by \citet{Gordon11} and Gordon et al., (in preparation).   These fluxes 
were measured within circular apertures of radius 2\fdg25  for the SMC and 3\fdg75 for the LMC.  For both galaxies, a sky annulus with a 0\fdg2 width was used to estimate and subtract the
background/foreground.   To within the uncertainties, the integrated DIRBE fluxes at 3.5~\um\  from Israel et al. match the \Spitzer\ fluxes at 3.6~\um\ from Gordon et al. The \Spitzer\ fluxes are provided in Table \ref{tab:results}.%We apply a statistical correction for foreground contamination
%of Milky Way stars as provided by TRILEGAL \citet{Girardi05}.

The spectral energy distributions of the SMC and LMC are shown in Figure \ref{fig:AGB}.  Also overlaid are dust model predictions from \citet{Bot10}. While the stellar flux is declining with wavelength, the warm dust and PAH components are increasing.    

We compare these integrated galaxy fluxes gathered from the literature to the fluxes from the TP-AGB and
RSG stars as measured from 2MASS and {\it Spitzer}
photometry.  The 2MASS $J$ and $K_{\rm S}$-bands are similar to
the COBE/DIRBE 1.25 and 2.2~\micron\ filters, however the DIRBE filters are 0.05 \um\ (15\%) wider at the long wavelength end. Similarly, the COBE/DIRBE 3.5~\micron\ filter is slightly bluer and about
25\% wider than the \Spitzer\ 3.6~\um\ filter.  These slight filter miss-matches may increase the uncertainties on the measured AGB and RSG flux fractions by modest amounts.  There are no filter mis-matches for the $3.6 - 24$~\um\ flux ratios, as all measurements are made from the same \Spitzer\ images.   

%The DIRBE 4.9-\micron\ band overlaps both the IRAC
%4.5~\micron\ and 5.8~\micron\ filters, but is a closer match to the
%IRAC 4.5-\micron\ band. However, the DIRBE band is considerably more narrow,
%covering 4.6--5.2~\micron, while the IRAC filter covers
%4--5~\micron. 
%The flux from AGB and RSG stars in the
%4.5-\micron\ IRAC band is thus overestimated compared to the global
%flux of each galaxy in the DIRBE 4.9~\micron\ band.

%The 24-\micron\ {\it Spitzer}/MIPS band covers approximately
%21--26~\micron. The COBE/DIRBE 25~\micron\ band is considerably wider
%(17--26~\micron). The stellar flux as measured with the MIPS 24-\micron\ band
%will thus be underestimated compared to the DIRBE band. 

\subsection{Uncertainties}

In the following section we report the fractional flux contribution of TP-AGB and RHeB stars to the Magellanic Clouds as given by:

\begin{equation}
\mathrm{Flux \; Fraction} = \frac{\mathrm{Stellar \;Flux - Foreground \; Stellar\; Flux}}{\mathrm{Integrated \;Galaxy \;Flux}}.
\end{equation}

  The uncertainties for these measurements are constructed from the following components.  First, we assume the published uncertainties for the integrated fluxes of the Magellanic Clouds at each wavelength.  Second, we compile the photometric uncertainty in the integrated stellar flux from the photometric uncertainty of each star added in quadrature.  Third, we estimate the uncertainty on the foreground correction by taking the Poisson uncertainty on the numbers of foreground stars in each classification multiplied by the typical flux of those stars.  We propagate these uncertainties through Equation 1. Fourth, we include an estimate of the uncertainty introduced by AGB stars that are fainter than the TRGB as described in section 2.1.

Finally, we have quantified how much of the TP-AGB and RSG population
  we have missed due to limited spatial coverage within the apertures
  we used to measure each galaxy's integrated flux.  In both galaxies,
  the TP-AGB and RSG stars follow a spherical distribution centered on
  the Bars (e.g., Blum et al. 2006; Boyer et al. 2011). For the SMC,
  the SAGE footprint includes an additional strip of coverage that
  extends $>$6 degrees along the SMC Tail, and for the LMC, the SAGE
  coverage stretches along the Disk, $>$4.5 degrees from the Bar's
  center.  We can use this extended coverage to measure the radial
  profile of AGB stars in the Bars by counting the number of AGB stars
  in concentric ellipses centered on the Bars.  We find that the AGB
  population is rather tightly concentrated, dropping to very small
  numbers by about 2 degrees (along the minor axes) from the
  population centers.  By extrapolating these radial profiles in all
  directions, we estimate that we have missed about 500 SMC AGB stars
  and 900 LMC AGB stars (10 SMC RSGs and 20 LMC RSGs) by restricting
  our apertures to 2.25 and 3.75 degrees for the SMC and LMC,
  respectively. Assuming that the missed stars have a similar flux distribution as the measured TP-AGB/RSG population, this
  amounts to an extra $<$1\% uncertainty on the integrated AGB (and
  RSG) flux, which we have included in the upper uncertainties in Table 1 and Figure 2.

\section{Results: The Multi-Wavelength Contribution of TP-AGB and RSG Stars to the Infrared Flux of the Magellanic Clouds}

\begin{figure*}
\centering
\plotone{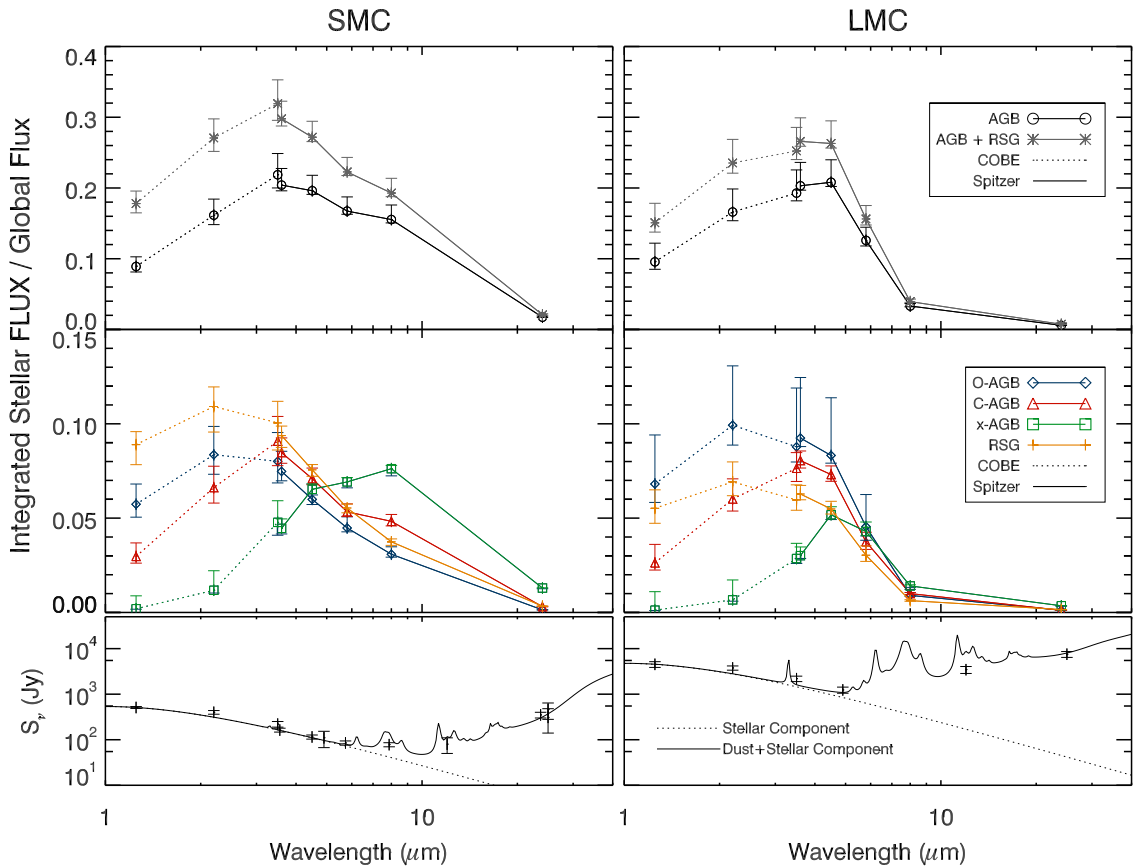}
\caption{\label{fig:AGB} The fractional contribution from TP-AGB stars (open symbols) and RSG stars (plus symbols) to the integrated near- and mid-IR fluxes of the SMC and LMC (top and middle panels).  The contributions divided by sub-species are given in the middle panels, while the total contributions from all TP-AGB stars (circles) and TP-AGB $+$ RSG (asterix) are shown in the top panels. The bottom panel shows the integrated SEDs of the LMC and SMC and a model of the dust emission from \citet{Bot10}.  While the stellar flux declines with wavelength (dotted line), the dust emission increases. The flux contribution of TP-AGB stars peaks at $\sim3-4$ \um.  The contribution from RSG stars peaks at shorter wavelengths (roughly 2.2~\um).   Combined, these evolved stars contribute 32\% of the SMC light at $3-4$ \um\ and 19\% at 8~\um. The flux contributions of the TP-AGB and RSG at 24 \um\ are small, $< 3$\%, in both galaxies.   }
\end{figure*}

Figure \ref{fig:AGB} shows the fraction of the integrated galaxy light
produced by TP-AGB (open symbols) and RSG (plus symbols) stars in the
SMC and LMC as a function of wavelength from 1--24~\micron.  These results are
also presented in Table 1. At 3.6~\micron, these rare stars
contribute over 30\% of the SMC flux even though they represent a
negligible fraction of the stellar mass of the galaxy 
(numbers of stars of each type are also shown in Table 1). In the LMC,
they contribute more than a quarter of the 3.6~\um\ light.  
At 24~\micron, the stellar contribution to the total luminosity is $<$3\% for both galaxies.

The flux contribution from RSG stars, which are bluer than AGB stars,
peaks near $\sim2$ \micron.  In the SMC, they are responsible for over 10\%
of the $K$-band light.  In the LMC, the fraction is  lower,
7\%. Part of this difference is no doubt due to the detailed star
formation histories of these two galaxies over the last 0.1 Gyr, with
the SMC having a higher specific star formation rate over that period
\citep[e.g.][]{Harris04,Harris09}.  

%{\bf I think HZ 2007 find a peak in the SFH at 60 Myr, corresponding
 % to $\gtrsim$8 solar masses.  If the LMC doesn't have something similar,
 % that could also help explain things. But contamination of the RSG
 % sample is an important issue.}

The flux contribution from TP-AGB stars peaks in the {\it Spitzer} IRAC
bands between 3 and 4~\micron, reaching $15-20$\% for both the LMC and
SMC.  The similarity of the results for the two galaxies suggests
that over the longer time scale of $0.1 - 2$ Gyr, these galaxies have
had similar star formation histories, as has been inferred by studies
of their CMDs \citep{Harris04,Harris09}.

%These results compliment our previous study of the contribution of TP-AGB and RSG stars to the 1.6 \micron\ fluxes of 23 nearby dwarfs and spirals \citep{Melbourne12}.  The TP-AGB and RSG contribution to the IR-luminosities of the SMC and LMC at 1.6 \micron\ is at the high end of those found for the larger local sample, but is reasonable given the high specific star formation rates of the Clouds.  

Figure \ref{fig:AGB} also shows the TP-AGB populations sub-divided into the 3 categories previously discussed: 
O-AGB, C-AGB, and x-AGB.  Despite the fact that the O-AGB population outnumbers the C-AGB
population by a factor of 1.8 in the SMC and 2.4 in the LMC, the
contribution to the global flux of the O-AGB and C-AGB populations are
similar at $\lambda > 3$~\micron, with each contributing roughly equal
parts to the total TP-AGB flux.  However, the contribution from O-AGB
stars peaks at shorter wavelengths (near 2.2~\micron) than the C-AGB star
contribution (near 3.6~\micron). In contrast, the x-AGB stars account for
less than 1\% of the global flux at $\lambda < 2$~\micron, but exceed the
flux contribution of the other two TP-AGB types  at $\lambda > 4$~\micron.
This is true even though the x-AGB population is tiny by numbers,
comprising $<$4\% of the total number of TP-AGB stars in both galaxies.

\section{Discussion}

We present, for the first time, the near- through mid-IR  flux contributions of 
complete samples of TP-AGB and RSG stars to the 
integrated luminosities of the Magellanic Clouds.  We find that 
these stars can contribute non-negligible flux fractions from 1 to 8~\um\ (e.g., $15-30$\%), and smaller fractions at 24~\um (e.g., $< 3$\%).  Below we compare these results to previous work and discuss the implications for estimating stellar masses and star formation rates of galaxies.

\subsection{The Importance of TP-AGB and RSG Stars at Near-IR Wavelengths}
Rest-frame near-IR fluxes typically anchor the stellar mass estimates
of galaxies \citep[e.g.,][]{Maraston06, Conroy09, Zibetti09,Eskew12}.
At these wavelengths, dust obscuration is significantly reduced
compared with optical light, and a galaxy's near-IR luminosity is much
less affected by the short-lived O- and B-type stars produced by
current star formation.  Instead, the near-IR is typically thought to
be dominated by older, easily modeled RGB stars.

However, we have shown that TP-AGB and massive RSG stars can also account for large fractions of the near-IR light (e.g. $> 30$\% at 3.6~\um\ for the SMC), even though they contribute essentially nothing to the total stellar masses of galaxies. A galaxy with the same 3.6~\micron\ luminosity as the SMC but comprised of entirely old stellar populations (e.g., $> 3$ Gyr) must be nearly twice as massive as the SMC to make up for the missing stellar luminosity provided by the TP-AGB and RSG at that wavelength. Thus, a galaxy's flux at 3--4~\micron\ (and hence its M/L ratio) is highly sensitive to the SFH.

The large TP-AGB and RSG near-IR flux contributions we measure for the Magellanic Clouds fit well with the contributions reported in \citet{Melbourne12a} for a sample of 23 nearby dwarfs and small spirals, which ranged from $5 - 40$\% at 1.6~\um. Similarly, our results are well matched to the predictions of \citet{Bruzual11}, which discussed the  contribution of AGB stars to the $K$-band flux of their stellar population synthesis models.  Their results for a galaxy with $Z = 0.008$ and a declining SFR with an e-folding time of 1~Gyr are consistent with what we see in the Magellanic Clouds.
These large flux contributions suggest that galaxies with even modest populations of TP-AGB and RSG stars will have very different near-IR M/L ratios than galaxies lacking these populations, making near-IR based stellar mass estimates difficult.

\begin{figure*}
\centering
\includegraphics[scale=0.8]{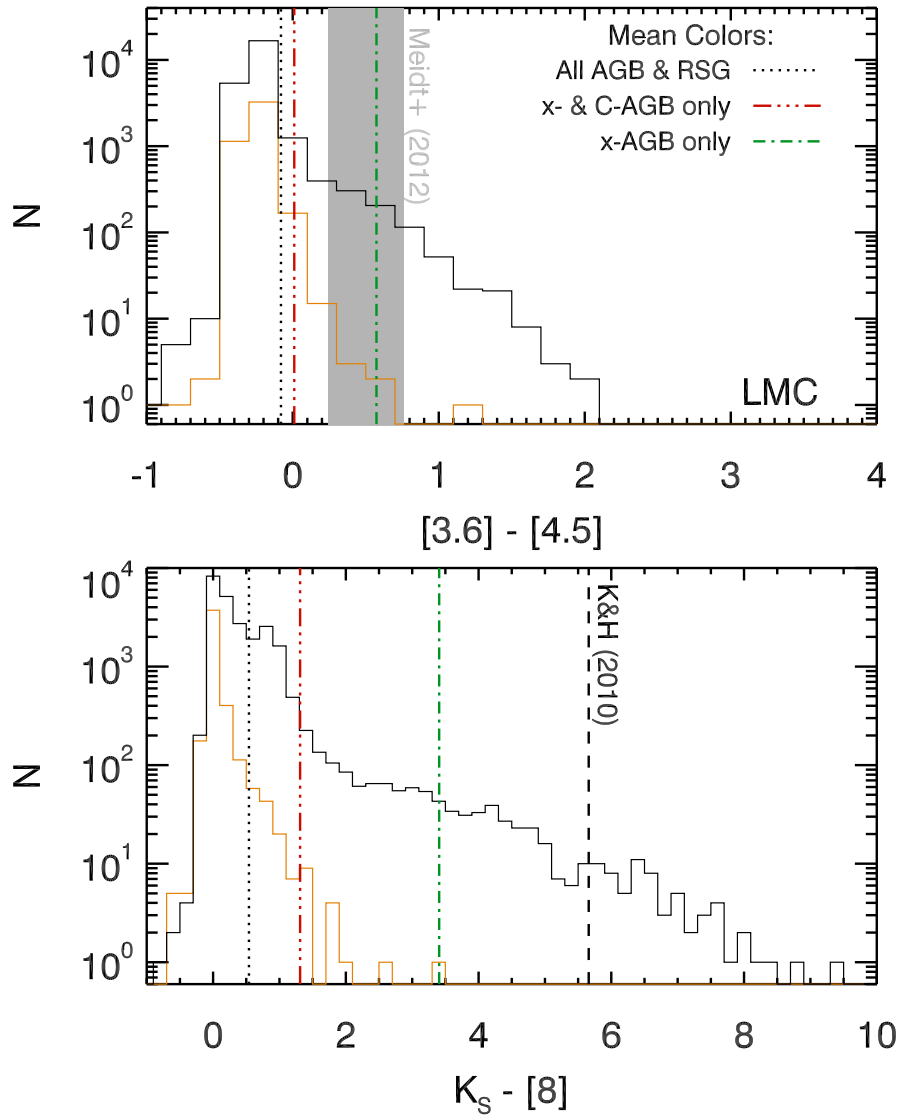}
\caption{\label{fig:hist}Histograms of the $[3.6]-[4.5]$ (top) and $K_{\rm S} - [8]$ (bottom) colors of the AGB (black) and RSG (yellow) stars in the LMC. Also shown are various color selections used in the literature to select AGB populations. The top panel shows the $[3.6]-[4.5]$ color range used by \citet{meidt12} (shaded region) to isolate AGB and RSG stars, which roughly matches the mean selection used in \citet{Gerke12}. Both of these selections are significantly redder than the mean AGB and RSG colors in the LMC (dotted line; and the SMC, see Boyer et al. 2011).  The lower panel shows the typical carbon star color used in the \citet{Kelson10} models (dashed line), which is significantly redder than the mean color for the x-AGB stars (green dash-dot line) and over 4 magnitudes redder than the mean carbon star color (red dash-dot-dot-dot line).}
\end{figure*}

%At shorter wavelengths, the TP-AGB and 
%RSG stars together contribute larger fractions, $\approx$30\% of the 
%flux at 3.6~\micron\ in the SMC and 25\% in the LMC.   
%At 1.2~\micron, these stars, are responsible for 15--25\% of the global flux.
%\citet{Bruzual09} discusses the  Likewise, t  %They also find that the AGB stars can contribute
%20\%--60\% of the rest-frame $K$-band flux at redshifts between 1 and
%7.5, so it is clear that the AGB stars cannot be neglected in
%high-redshift galaxies.  

However, some authors find significantly smaller near-IR flux contributions from TP-AGB stars. For instance, \citet{Gerke12}  calculate a smaller 3.6 \um\ flux contribution from the TP-AGB  in the Magellanic clouds, only 5\% of the integrated light. They selected AGB stars using the $[3.6]-[4.5]$ color, however, they selected a CMD region that is comprised predominantly of x-AGB stars. Thus they miss the bulk of the TP-AGB altogether.  The top panel of Figure \ref{fig:hist}  shows the distribution of  $[3.6]-[4.5]$ colors of TP-AGB stars in the LMC.  The \citet{Gerke12} sample has a mean color similar to the x-AGB alone (green dot-dashed line), which are significantly redder than the overall sample (dotted line).   If we compare only the x-AGB stars in our sample to their result, we find a similarly small contribution of $\sim3$\%. Including the full TP-AGB sample results in a much higher flux contribution at 3.6~\um, as discussed above.  

\citet{Meidt12a} shows a similar result to \citet{Gerke12}, only they used unresolved \Spitzer\ 3.6 and 4.5 ~\um\ observations of galaxies in the \Spitzer\ Survey of Stellar Structure in Galaxies \citep[S$^4$G][]{Sheth10}.  Their work spatially separates (on a pixel-by-pixel basis) older stellar populations from warm dust, PAH emission, and evolved stars, based on the colors of regions in the \Spitzer\ data.  Thus they are able to nicely constrain the warm dust and PAH emission to roughly $5-15$\% of the 3.6~\um\ flux in a sample of 6 nearby galaxies.  Likewise they find that pixels dominated by evolved AGB and RSG stars account for only another roughly 5\% of the 3.6 \um\ light.  However, as with the \citet{Gerke12} result, their assumed AGB color of $[3.6]-[4.5]=0.24 - 0.76$ (grey shaded region of Figure \ref{fig:hist}) is significantly redder than the bulk of the TP-AGB stars in the LMC which have a mean color of $[3.6]-[4.5]=0.0$ (red dot-dot-dot-dashed line).  In fact, the mean $[3.6]-[4.5]$ color for TP-AGB and RSG  in the LMC is $[3.6]-[4.5]=-0.07$ (dotted line), exactly the color \citet{Meidt12a} assign to old stars. Unfortunately, that means that \citet{Meidt12a} has only selected the reddest TP-AGB stars and, as a result, they are likely under-estimating the contribution of TP-AGB stars at 3.6~\um.  %However, accounting for the redder x-AGB is still extremely helpful.

\begin{figure}
\centering
\plotone{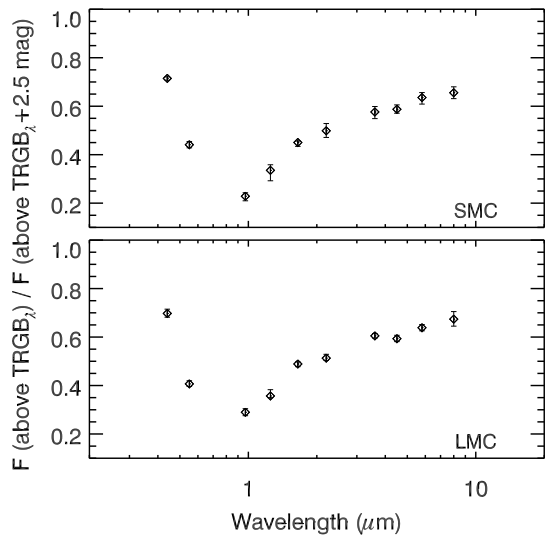}
\caption{\label{fig:bestlam} The contribution of luminous stars (with fluxes $>$ than the TRGB) to the total point-source flux (measured to 2.5 mags below the the TRGB) as a function of wavelength.  At the blue end the O and B stars contribute significantly to the point-source flux.  At the red end, RSG and TP-AGB stars contribute significantly to the total.  As all of these stars are short lived, they will result in a rapidly changing M/L ratio at these wavelengths. The impact of luminous short-lived stars on the stellar mass estimates of galaxies can be minimized by using the M/L ratio at $0.8 - 1.0$~\um, where this plot shows a minimum.}
\end{figure}

The results we have presented here suggest that the near-IR may not be optimized for stellar mass estimates of galaxies.   The 3.6~\micron\ band, in particular, is the most problematic for the effects of  short-lived TP-AGB and RSG stars, which can impose a stochastic variability on the near-IR M/L ratio of stellar populations \citep[e.g.,][]{Melbourne12a}.  However, it should be possible to minimize the effects of the TP-AGB and RSG stars on estimates of the near-IR M/L ratios of galaxies by using bluer wavebands.   The TP-AGB and RSGs flux contribution at 1~\um\ is roughly half that of  the 3.6-\micron\ flux contribution.  Of course, bluer wavelengths will be more impacted by extinction from dust and emission from hot O and B stars, but the $I$- or $J$-band may prove to be the ideal compromise.  To test this hypothesis we compare the point-source flux above the TRGB to the total point-source flux (which is measurable to 2.5 mags below the TRGB)  as a function of wavelength from $0.5 - 3.6$~\um. Figure \ref{fig:bestlam} displays the results, and shows that indeed there is a minimum flux contribution from luminous short-lived stars as a function of wavelength. The minimum occurs between 0.8 and 1 \um.  At shorter wavelengths, massive main sequence O and B stars contribute large fractions of light.  At longer wavelengths, the TP-AGB and RSG stars contribute large fractions.   This plot suggests that rest-frame  $I-$ through $J-$band will indeed provide M/L ratios that are the most stable against stellar population history and the influence of these short-lived luminous stars.

\subsection{The Importance of TP-AGB and RSG Stars at Mid-IR Wavelengths}

Because mid-IR light is produced by warm dust and PAH emission in star forming regions, mid-IR luminosities of galaxies are often used as proxies for their current star formation rates. However, we have shown that the TP-AGB and RSG stars can also contribute to the mid-IR fluxes of galaxies, with the bulk of the mid-IR contribution coming from the rare but extremely dusty x-AGB stars.  In the SMC, the contribution of TP-AGB and RSG stars at 8~\micron\ is $\approx$19\%! If these x-AGB stars are neglected in the accounting of the mid-IR light, an SMC SFR based on 8~\um\ light alone will be biased high. 

This result has impacts on the interpretation of thousands of mid-IR based star formation rates, for example at $z=2$, \Spitzer\ 24~\um\ imaging traces rest-frame 8~\um, where the TP-AGB is potentially an important flux contributor. However, the size of the effect may be small in practice.  For instance, in the LMC, the TP-AGB + RSG 8~\um\ contribution is $< 5$\%. Compared with the SMC, the LMC has a significantly larger contribution from warm dust and  8~\um\ PAH  emission \citep[][and Figure \ref{fig:AGB} here]{Bot10}.  Samples of the extremely dusty $z=2$ galaxies detected by \Spitzer\ have been shown to have large PAH flux contributions \citep[e.g.][]{Desai09}, so the correction for TP-AGB stars may be small in the high-$z$ luminous infrared galaxies.  However,  it is possible that TP-AGB stars artificially elevate the the rest-frame 8~\um\ estimated SFRs of galaxies, especially for lower-metallicity systems like the SMC where (1) PAH emission is limited, (2) C-AGB stars are more common, and (3) star formation rates are likely smaller.

We are not the first to point out that the TP-AGB may be important at mid-IR wavelengths.  For instance, \citet{Kelson10}, and \citet{Chisari12} recently proposed TP-AGB stars as the primary source of mid-IR flux in galaxies, and may account for the rise of luminous infrared galaxies with redshift. Similarly, \citet{Verley09} suggested dusty shells surrounding AGB stars could account for the bulk of diffuse 24~\um\ emission in the outskirts of M33.  While our results show the importance of TP-AGB stars at mid-IR wavelengths, these earlier works likely overemphasize the TP-AGB contribution, especially at 24 \um, where we show that TP-AGB stars contribute less than 3\% of the Magellanic Cloud flux.

The large mid-IR TP-AGB contributions predicted by these previous works (e.g. $> 50$\% of the 8 and 24 \um\ flux) likely arose from their adopted mid-IR colors and fluxes of C-AGB stars. \citet{Kelson10} and \citet{Chisari12} assume that C-rich AGB stars have an average $K-[8.8]$ color of $5.66 \pm 1.16$~mag, derived from Galactic carbon stars observed in the mid-IR by \citet{Guandalini06}. \citet{Verley09} assume an average 24~\micron\ flux derived from carbon stars observed with {\it Spitzer} in the Magellanic Clouds by \citet{Groenewegen07}.  However, both the \citet{Guandalini06} and \citet{Groenewegen07} samples are heavily biased towards the reddest x-AGB stars and completely exclude the bulk of the C-AGB population, which is much bluer.  This issue is demonstrated in the bottom panel of Figure \ref{fig:hist} which plots histograms of the $K_{\rm S}-[8]$ color of the TP-AGB stars in the LMC. The color adopted by \citet{Kelson10} models is shown as a dashed line and is significantly redder than the bulk of the LMC C-AGB stars. When including the complete samples of C-rich AGB stars (C-AGB plus x-AGB, red and green distributions respectively) from the SAGE survey, we find $\langle K_{\rm S} - [8] \rangle_{\rm x + C} = 1.38$~mag in the SMC and $1.31$~mag in the LMC (dot-dot-dot-dashed line in Figure \ref{fig:hist}).  This is over 4 magnitudes bluer than the Kelson \& Holden models \citep[also used in ][]{Chisari12}.  A similarly sized discrepancy is obtained when comparing the SAGE AGB sample to the adopted 24~\um\ flux for AGB stars in Verley et al. These revised colors would result in a much smaller contribution of TP-AGB stars at mid-IR wavelengths, as is seen Figure~\ref{fig:AGB} for the Magellanic Clouds.

While some x-AGB stars in the SAGE Magellanic Cloud sample do show colors as red as those
used in the Kelson \& Holden models, together they have an average color
that is still 2 magnitudes bluer than the models, $\langle K_{\rm S} - [8] \rangle_{\rm x,\ SMC} =
3.25$~mag and $\langle K_{\rm S} - [8] \rangle_{\rm x,\ LMC} =
3.41$~mag (dot-dashed line in top panel of Figure \ref{fig:hist}).  
In addition, these x-AGB stars contribute almost nothing 
to the K-band light of the Magellanic clouds (1\% or less). The bluer O-AGB  and C-AGB stars produce roughly 20 times
the $K$-band light of the x-AGB stars. The Kelson \& Holden models rely on the $K$-band flux 
to estimate the AGB contribution at longer wavelengths, and they assume that the x-ABG stars 
are contributing nearly 50\% of the $K$-band light.  Because of these assumptions, they are likely significantly over-estimating the AGB flux contribution at longer wavelengths.

\section{Conclusions}
The contribution of TP-AGB and RSG stars to the integrated fluxes of the Magellanic Clouds, varies with wavelength.  We find that their combined contribution peaks between $3 - 4$~\um, where they contribute 32\% of the SMC flux and $\sim25$\% of the LMC flux.  At 1.2~\um, the contributions are smaller, $\sim18$\% and $\sim15$\% for the SMC and LMC.   At longer wavelengths the contributions also decline, but still constitute 19\% of the 8~\um\ light of the SMC (only 4\% for the LMC where warm dust and PAH emission are stronger). The bulk of this 8~\um\ stellar flux is from the most dust-enshrouded AGB stars, the so-called x-AGB stars, which are predominantly carbon stars near the end of the AGB phase.  At 24~\um\ the AGB star contribution to the total flux is less than 3\% for each galaxy.

The flux contributions of the TP-AGB and RSG are significant at 3.6~\um.  Thus, these stars have an impact on the M/L ratios of the Magellanic Clouds at near-IR wavelengths.  Because these stars are short-lived we can expect that the near-IR M/L ratios will also vary significantly over time.  To minimize the impact of short lived TP-AGB and RSG stars on stellar mass estimates of galaxies, shorter wavelength observations are preferred.  However, at optical wavelengths, short-lived, super-luminous O and B stars play a similar role as the AGB and RSG stars in the near-IR.  We find that the optimal wavelengths for minimizing the impact of luminous short-lived stars on the M/L ratio of galaxies is between 0.8 and 1~\um.   

The non-negligible flux from the TP-AGB at 8~\um\ suggests that star formation rates based on rest-frame 8~\um\ flux (e.g. \Spitzer\ 24~\um\ imaging at $z=2$) may be biassed slightly high.  However, this effect is most important for systems with little PAH emission (like the SMC).  For systems with significant warm dust and PAH emission, the TP-AGB contribution at 8~\um\ appears to be small, thus for the LMC the contribution is $<$4\%.

\acknowledgments
Special thanks to Karl Gordon for providing integrated fluxes of the LMC in the \Spitzer\ bands, to Caroline Bot for sharing the SED and model fits of the Magellanic Clouds. Thanks to Gustavo Bruzual for sharing updated SSP models of the Magellanic Clouds. We also want to thank Barry Madore, Mark Seibert, and Wendy Freedman for inspiring efforts to study the impact of AGB stars on mid-infrared galaxy fluxes. This work is based in part on observations made with the {\it Spitzer Space Telescope}, which is operated by the Jet Propulsion Laboratory, California Institute of Technology under a contract with NASA.

\bibliographystyle{/Users/jmel/bib/apj}

\begin{thebibliography}{57}
\expandafter\ifx\csname natexlab\endcsname\relax\def\natexlab#1{#1}\fi

\bibitem[{{Bell} \& {de Jong}(2001)}]{Bell01}
{Bell}, E.~F., \& {de Jong}, R.~S. 2001, \apj, 550, 212

\bibitem[{{Bertelli} {et~al.}(2008){Bertelli}, {Girardi}, {Marigo}, \&
  {Nasi}}]{Bertelli08}
{Bertelli}, G., {Girardi}, L., {Marigo}, P., \& {Nasi}, E. 2008, \aap, 484, 815

\bibitem[{{Bertelli} {et~al.}(2009){Bertelli}, {Nasi}, {Girardi}, \&
  {Marigo}}]{Bertelli09}
{Bertelli}, G., {Nasi}, E., {Girardi}, L., \& {Marigo}, P. 2009, \aap, 508, 355

\bibitem[{{Blum} {et~al.}(2006){Blum}, {Mould}, {Olsen}, {Frogel}, {Werner},
  {Meixner}, {Markwick-Kemper}, {Indebetouw}, {Whitney}, {Meade}, {Babler},
  {Churchwell}, {Gordon}, {Engelbracht}, {For}, {Misselt}, {Vijh}, {Leitherer},
  {Volk}, {Points}, {Reach}, {Hora}, {Bernard}, {Boulanger}, {Bracker},
  {Cohen}, {Fukui}, {Gallagher}, {Gorjian}, {Harris}, {Kelly}, {Kawamura},
  {Latter}, {Madden}, {Mizuno}, {Mizuno}, {Nota}, {Oey}, {Onishi}, {Paladini},
  {Panagia}, {Perez-Gonzalez}, {Shibai}, {Sato}, {Smith}, {Staveley-Smith},
  {Tielens}, {Ueta}, {Van Dyk}, \& {Zaritsky}}]{Blum06}
{Blum}, R.~D., {Mould}, J.~R., {Olsen}, K.~A., {Frogel}, J.~A., {Werner}, M.,
  {Meixner}, M., {Markwick-Kemper}, F., {Indebetouw}, R., {Whitney}, B.,
  {Meade}, M., {Babler}, B., {Churchwell}, E.~B., {Gordon}, K., {Engelbracht},
  C., {For}, B., {Misselt}, K., {Vijh}, U., {Leitherer}, C., {Volk}, K.,
  {Points}, S., {Reach}, W., {Hora}, J.~L., {Bernard}, J., {Boulanger}, F.,
  {Bracker}, S., {Cohen}, M., {Fukui}, Y., {Gallagher}, J., {Gorjian}, V.,
  {Harris}, J., {Kelly}, D., {Kawamura}, A., {Latter}, W.~B., {Madden}, S.,
  {Mizuno}, A., {Mizuno}, N., {Nota}, A., {Oey}, M.~S., {Onishi}, T.,
  {Paladini}, R., {Panagia}, N., {Perez-Gonzalez}, P., {Shibai}, H., {Sato},
  S., {Smith}, L., {Staveley-Smith}, L., {Tielens}, A.~G.~G.~M., {Ueta}, T.,
  {Van Dyk}, S., \& {Zaritsky}, D. 2006, \aj, 132, 2034

\bibitem[{{Bot} {et~al.}(2010){Bot}, {Ysard}, {Paradis}, {Bernard}, {Lagache},
  {Israel}, \& {Wall}}]{Bot10}
{Bot}, C., {Ysard}, N., {Paradis}, D., {Bernard}, J.~P., {Lagache}, G.,
  {Israel}, F.~P., \& {Wall}, W.~F. 2010, \aap, 523, A20

\bibitem[{{Bowen}(1988)}]{Bowen88}
{Bowen}, G.~H. 1988, \apj, 329, 299

\bibitem[{{Boyer} {et~al.}(2009){Boyer}, {Skillman}, {van Loon}, {Gehrz}, \&
  {Woodward}}]{Boyer09}
{Boyer}, M.~L., {Skillman}, E.~D., {van Loon}, J.~{\relax Th}., {Gehrz}, R.~D.,
  \& {Woodward}, C.~E. 2009, \apj, 697, 1993

\bibitem[{{Boyer} {et~al.}(2011){Boyer}, {Srinivasan}, {van Loon}, {McDonald},
  {Meixner}, {Zaritsky}, {Gordon}, {Kemper}, {Babler}, {Block}, {Bracker},
  {Engelbracht}, {Hora}, {Indebetouw}, {Meade}, {Misselt}, {Robitaille},
  {Sewi{\l}o}, {Shiao}, \& {Whitney}}]{Boyer11}
{Boyer}, M.~L., {Srinivasan}, S., {van Loon}, J.~{\relax Th}., {McDonald}, I.,
  {Meixner}, M., {Zaritsky}, D., {Gordon}, K.~D., {Kemper}, F., {Babler}, B.,
  {Block}, M., {Bracker}, S., {Engelbracht}, C.~W., {Hora}, J., {Indebetouw},
  R., {Meade}, M., {Misselt}, K., {Robitaille}, T., {Sewi{\l}o}, M., {Shiao},
  B., \& {Whitney}, B. 2011, \aj, 142, 103

\bibitem[{{Bruzual}(2011)}]{Bruzual11}
{Bruzual}, A. G. 2011, in Revista Mexicana de Astronomia y Astrofisica, vol.
  27, Vol.~40, Revista Mexicana de Astronomia y Astrofisica Conference Series,
  36--41

\bibitem[{{Chary} \& {Elbaz}(2001)}]{Chary01}
{Chary}, R., \& {Elbaz}, D. 2001, \apj, 556, 562

\bibitem[{{Chisari} \& {Kelson}(2012)}]{Chisari12}
{Chisari}, N.~E., \& {Kelson}, D.~D. 2012, \apj, 753, 94

\bibitem[{{Cioni} {et~al.}(2006){Cioni}, {Girardi}, {Marigo}, \&
  {Habing}}]{Cioni06}
{Cioni}, M., {Girardi}, L., {Marigo}, P., \& {Habing}, H.~J. 2006, \aap, 448,
  77

\bibitem[{{Conroy} {et~al.}(2009){Conroy}, {Gunn}, \& {White}}]{Conroy09}
{Conroy}, C., {Gunn}, J.~E., \& {White}, M. 2009, \apj, 699, 486

\bibitem[{{Dalcanton} {et~al.}(2012){Dalcanton}, {Williams}, {Melbourne},
  {Girardi}, {Dolphin}, {Rosenfield}, {Boyer}, {de Jong}, {Gilbert}, {Marigo},
  {Olsen}, {Seth}, \& {Skillman}}]{Dalcanton12a}
{Dalcanton}, J.~J., {Williams}, B.~F., {Melbourne}, J.~L., {Girardi}, L.,
  {Dolphin}, A., {Rosenfield}, P.~A., {Boyer}, M.~L., {de Jong}, R.~S.,
  {Gilbert}, K., {Marigo}, P., {Olsen}, K., {Seth}, A.~C., \& {Skillman}, E.
  2012, \apjs, 198, 6

\bibitem[{{Dalcanton et al.}(2011)}]{Dalcanton11}
{Dalcanton et al.} 2011, \apjs, in preparation

\bibitem[{{Desai} {et~al.}(2009){Desai}, {Soifer}, {Dey}, {Le Floc'h}, {Armus},
  {Brand}, {Brown}, {Brodwin}, {Jannuzi}, {Houck}, {Weedman}, {Ashby},
  {Gonzalez}, {Huang}, {Smith}, {Teplitz}, {Willner}, \& {Melbourne}}]{Desai09}
{Desai}, V., {Soifer}, B.~T., {Dey}, A., {Le Floc'h}, E., {Armus}, L., {Brand},
  K., {Brown}, M.~J.~I., {Brodwin}, M., {Jannuzi}, B.~T., {Houck}, J.~R.,
  {Weedman}, D.~W., {Ashby}, M.~L.~N., {Gonzalez}, A., {Huang}, J., {Smith},
  H.~A., {Teplitz}, H., {Willner}, S.~P., \& {Melbourne}, J. 2009, \apj, 700,
  1190

\bibitem[{{D{\'{\i}}az-Santos} {et~al.}(2008){D{\'{\i}}az-Santos},
  {Alonso-Herrero}, {Colina}, {Packham}, {Radomski}, \&
  {Telesco}}]{Diaz-Santos08}
{D{\'{\i}}az-Santos}, T., {Alonso-Herrero}, A., {Colina}, L., {Packham}, C.,
  {Radomski}, J.~T., \& {Telesco}, C.~M. 2008, \apj, 685, 211

\bibitem[{{Dohm-Palmer} \& {Skillman}(2002)}]{Dohm-Palmer02}
{Dohm-Palmer}, R.~C., \& {Skillman}, E.~D. 2002, \aj, 123, 1433

\bibitem[{{Elitzur} \& {Ivezi{\'c}}(2001)}]{Elitzur01}
{Elitzur}, M., \& {Ivezi{\'c}}, {\v Z}. 2001, \mnras, 327, 403

\bibitem[{{Eskew} {et~al.}(2012){Eskew}, {Zaritsky}, \& {Meidt}}]{Eskew12}
{Eskew}, M.~R., {Zaritsky}, D.~F., \& {Meidt}, S.~E. 2012, ArXiv e-prints

\bibitem[{{Fazio} {et~al.}(2004){Fazio}, {Hora}, {Allen}, {Ashby}, {Barmby},
  {Deutsch}, {Huang}, {Kleiner}, {Marengo}, {Megeath}, {Melnick}, {Pahre},
  {Patten}, {Polizotti}, {Smith}, {Taylor}, {Wang}, {Willner}, {Hoffmann},
  {Pipher}, {Forrest}, {McMurty}, {McCreight}, {McKelvey}, {McMurray}, {Koch},
  {Moseley}, {Arendt}, {Mentzell}, {Marx}, {Losch}, {Mayman}, {Eichhorn},
  {Krebs}, {Jhabvala}, {Gezari}, {Fixsen}, {Flores}, {Shakoorzadeh}, {Jungo},
  {Hakun}, {Workman}, {Karpati}, {Kichak}, {Whitley}, {Mann}, {Tollestrup},
  {Eisenhardt}, {Stern}, {Gorjian}, {Bhattacharya}, {Carey}, {Nelson},
  {Glaccum}, {Lacy}, {Lowrance}, {Laine}, {Reach}, {Stauffer}, {Surace},
  {Wilson}, {Wright}, {Hoffman}, {Domingo}, \& {Cohen}}]{Fazio04}
{Fazio}, G.~G., {Hora}, J.~L., {Allen}, L.~E., {Ashby}, M.~L.~N., {Barmby}, P.,
  {Deutsch}, L.~K., {Huang}, J.-S., {Kleiner}, S., {Marengo}, M., {Megeath},
  S.~T., {Melnick}, G.~J., {Pahre}, M.~A., {Patten}, B.~M., {Polizotti}, J.,
  {Smith}, H.~A., {Taylor}, R.~S., {Wang}, Z., {Willner}, S.~P., {Hoffmann},
  W.~F., {Pipher}, J.~L., {Forrest}, W.~J., {McMurty}, C.~W., {McCreight},
  C.~R., {McKelvey}, M.~E., {McMurray}, R.~E., {Koch}, D.~G., {Moseley}, S.~H.,
  {Arendt}, R.~G., {Mentzell}, J.~E., {Marx}, C.~T., {Losch}, P., {Mayman}, P.,
  {Eichhorn}, W., {Krebs}, D., {Jhabvala}, M., {Gezari}, D.~Y., {Fixsen},
  D.~J., {Flores}, J., {Shakoorzadeh}, K., {Jungo}, R., {Hakun}, C., {Workman},
  L., {Karpati}, G., {Kichak}, R., {Whitley}, R., {Mann}, S., {Tollestrup},
  E.~V., {Eisenhardt}, P., {Stern}, D., {Gorjian}, V., {Bhattacharya}, B.,
  {Carey}, S., {Nelson}, B.~O., {Glaccum}, W.~J., {Lacy}, M., {Lowrance},
  P.~J., {Laine}, S., {Reach}, W.~T., {Stauffer}, J.~A., {Surace}, J.~A.,
  {Wilson}, G., {Wright}, E.~L., {Hoffman}, A., {Domingo}, G., \& {Cohen}, M.
  2004, \apjs, 154, 10

\bibitem[{{Gerke} \& {Kochanek}(2012)}]{Gerke12}
{Gerke}, J.~R., \& {Kochanek}, C.~S. 2012, ArXiv e-prints

\bibitem[{{Girardi} {et~al.}(2005){Girardi}, {Groenewegen}, {Hatziminaoglou},
  \& {da Costa}}]{Girardi05}
{Girardi}, L., {Groenewegen}, M.~A.~T., {Hatziminaoglou}, E., \& {da Costa}, L.
  2005, \aap, 436, 895

\bibitem[{{Girardi} {et~al.}(2010){Girardi}, {Williams}, {Gilbert},
  {Rosenfield}, {Dalcanton}, {Marigo}, {Boyer}, {Dolphin}, {Weisz},
  {Melbourne}, {Olsen}, {Seth}, \& {Skillman}}]{Girardi10}
{Girardi}, L., {Williams}, B.~F., {Gilbert}, K.~M., {Rosenfield}, P.,
  {Dalcanton}, J.~J., {Marigo}, P., {Boyer}, M.~L., {Dolphin}, A., {Weisz},
  D.~R., {Melbourne}, J., {Olsen}, K.~A.~G., {Seth}, A.~C., \& {Skillman}, E.
  2010, \apj, 724, 1030

\bibitem[{{Gordon} {et~al.}(2011){Gordon}, {Meixner}, {Meade}, {Whitney},
  {Engelbracht}, {Bot}, {Boyer}, {Lawton}, {Sewi{\l}o}, {Babler}, {Bernard},
  {Bracker}, {Block}, {Blum}, {Bolatto}, {Bonanos}, {Harris}, {Hora},
  {Indebetouw}, {Misselt}, {Reach}, {Shiao}, {Tielens}, {Carlson},
  {Churchwell}, {Clayton}, {Chen}, {Cohen}, {Fukui}, {Gorjian}, {Hony},
  {Israel}, {Kawamura}, {Kemper}, {Leroy}, {Li}, {Madden}, {Marble},
  {McDonald}, {Mizuno}, {Mizuno}, {Muller}, {Oliveira}, {Olsen}, {Onishi},
  {Paladini}, {Paradis}, {Points}, {Robitaille}, {Rubin}, {Sandstrom}, {Sato},
  {Shibai}, {Simon}, {Smith}, {Srinivasan}, {Vijh}, {Van Dyk}, {van Loon}, \&
  {Zaritsky}}]{Gordon11}
{Gordon}, K.~D., {Meixner}, M., {Meade}, M.~R., {Whitney}, B., {Engelbracht},
  C., {Bot}, C., {Boyer}, M.~L., {Lawton}, B., {Sewi{\l}o}, M., {Babler}, B.,
  {Bernard}, J.-P., {Bracker}, S., {Block}, M., {Blum}, R., {Bolatto}, A.,
  {Bonanos}, A., {Harris}, J., {Hora}, J.~L., {Indebetouw}, R., {Misselt}, K.,
  {Reach}, W., {Shiao}, B., {Tielens}, X., {Carlson}, L., {Churchwell}, E.,
  {Clayton}, G.~C., {Chen}, C.-H.~R., {Cohen}, M., {Fukui}, Y., {Gorjian}, V.,
  {Hony}, S., {Israel}, F.~P., {Kawamura}, A., {Kemper}, F., {Leroy}, A., {Li},
  A., {Madden}, S., {Marble}, A.~R., {McDonald}, I., {Mizuno}, A., {Mizuno},
  N., {Muller}, E., {Oliveira}, J.~M., {Olsen}, K., {Onishi}, T., {Paladini},
  R., {Paradis}, D., {Points}, S., {Robitaille}, T., {Rubin}, D., {Sandstrom},
  K., {Sato}, S., {Shibai}, H., {Simon}, J.~D., {Smith}, L.~J., {Srinivasan},
  S., {Vijh}, U., {Van Dyk}, S., {van Loon}, J.~T., \& {Zaritsky}, D. 2011,
  \aj, 142, 102

\bibitem[{{Groenewegen} {et~al.}(2007){Groenewegen}, {Wood}, {Sloan},
  {Blommaert}, {Cioni}, {Feast}, {Hony}, {Matsuura}, {Menzies}, {Olivier},
  {Vanhollebeke}, {van Loon}, {Whitelock}, {Zijlstra}, {Habing}, \&
  {Lagadec}}]{Groenewegen07}
{Groenewegen}, M.~A.~T., {Wood}, P.~R., {Sloan}, G.~C., {Blommaert},
  J.~A.~D.~L., {Cioni}, M.-R.~L., {Feast}, M.~W., {Hony}, S., {Matsuura}, M.,
  {Menzies}, J.~W., {Olivier}, E.~A., {Vanhollebeke}, E., {van Loon}, J.~T.,
  {Whitelock}, P.~A., {Zijlstra}, A.~A., {Habing}, H.~J., \& {Lagadec}, E.
  2007, \mnras, 376, 313

\bibitem[{{Guandalini} {et~al.}(2006){Guandalini}, {Busso}, {Ciprini},
  {Silvestro}, \& {Persi}}]{Guandalini06}
{Guandalini}, R., {Busso}, M., {Ciprini}, S., {Silvestro}, G., \& {Persi}, P.
  2006, \aap, 445, 1069

\bibitem[{Harris \& Zaritsky(2004)}]{Harris04}
Harris, J., \& Zaritsky, D. 2004, The Astronomical Journal, 127, 1531

\bibitem[{Harris \& Zaritsky(2009)}]{Harris09}
---. 2009, The Astronomical Journal, 138, 1243

\bibitem[{{Hauser} {et~al.}(1998){Hauser}, {Arendt}, {Kelsall}, {Dwek},
  {Odegard}, {Weiland}, {Freudenreich}, {Reach}, {Silverberg}, {Moseley},
  {Pei}, {Lubin}, {Mather}, {Shafer}, {Smoot}, {Weiss}, {Wilkinson}, \&
  {Wright}}]{Hauser98}
{Hauser}, M.~G., {Arendt}, R.~G., {Kelsall}, T., {Dwek}, E., {Odegard}, N.,
  {Weiland}, J.~L., {Freudenreich}, H.~T., {Reach}, W.~T., {Silverberg}, R.~F.,
  {Moseley}, S.~H., {Pei}, Y.~C., {Lubin}, P., {Mather}, J.~C., {Shafer},
  R.~A., {Smoot}, G.~F., {Weiss}, R., {Wilkinson}, D.~T., \& {Wright}, E.~L.
  1998, \apj, 508, 25

\bibitem[{{Iben}(1983)}]{Iben83a}
{Iben}, Jr., I. 1983, \apjl, 275, L65

\bibitem[{{Ilbert} {et~al.}(2010){Ilbert}, {Salvato}, {Le Floc'h}, {Aussel},
  {Capak}, {McCracken}, {Mobasher}, {Kartaltepe}, {Scoville}, {Sanders},
  {Arnouts}, {Bundy}, {Cassata}, {Kneib}, {Koekemoer}, {Le F{\`e}vre}, {Lilly},
  {Surace}, {Taniguchi}, {Tasca}, {Thompson}, {Tresse}, {Zamojski}, {Zamorani},
  \& {Zucca}}]{Ilbert10}
{Ilbert}, O., {Salvato}, M., {Le Floc'h}, E., {Aussel}, H., {Capak}, P.,
  {McCracken}, H.~J., {Mobasher}, B., {Kartaltepe}, J., {Scoville}, N.,
  {Sanders}, D.~B., {Arnouts}, S., {Bundy}, K., {Cassata}, P., {Kneib}, J.,
  {Koekemoer}, A., {Le F{\`e}vre}, O., {Lilly}, S., {Surace}, J., {Taniguchi},
  Y., {Tasca}, L., {Thompson}, D., {Tresse}, L., {Zamojski}, M., {Zamorani},
  G., \& {Zucca}, E. 2010, \apj, 709, 644

\bibitem[{{Israel} {et~al.}(2010){Israel}, {Wall}, {Raban}, {Reach}, {Bot},
  {Oonk}, {Ysard}, \& {Bernard}}]{Israel10}
{Israel}, F.~P., {Wall}, W.~F., {Raban}, D., {Reach}, W.~T., {Bot}, C., {Oonk},
  J.~B.~R., {Ysard}, N., \& {Bernard}, J.~P. 2010, \aap, 519, A67

\bibitem[{{Kelson} \& {Holden}(2010)}]{Kelson10}
{Kelson}, D.~D., \& {Holden}, B.~P. 2010, \apjl, 713, L28

\bibitem[{{Maraston} {et~al.}(2006){Maraston}, {Daddi}, {Renzini}, {Cimatti},
  {Dickinson}, {Papovich}, {Pasquali}, \& {Pirzkal}}]{Maraston06}
{Maraston}, C., {Daddi}, E., {Renzini}, A., {Cimatti}, A., {Dickinson}, M.,
  {Papovich}, C., {Pasquali}, A., \& {Pirzkal}, N. 2006, \apj, 652, 85

\bibitem[Marigo \& Girardi(2007)]{Marigo07} Marigo, P., \& Girardi, L.\ 2007, \aap, 469, 239 

\bibitem[{{Marigo} {et~al.}(2008){Marigo}, {Girardi}, {Bressan}, {Groenewegen},
  {Silva}, \& {Granato}}]{Marigo08}
{Marigo}, P., {Girardi}, L., {Bressan}, A., {Groenewegen}, M.~A.~T., {Silva},
  L., \& {Granato}, G.~L. 2008, \aap, 482, 883

\bibitem[{{Mattsson} {et~al.}(2008){Mattsson}, {Wahlin}, {H{\"o}fner}, \&
  {Eriksson}}]{Mattsson08}
{Mattsson}, L., {Wahlin}, R., {H{\"o}fner}, S., \& {Eriksson}, K. 2008, \aap,
  484, L5

\bibitem[{{McQuinn} {et~al.}(2011){McQuinn}, {Skillman}, {Dalcanton},
  {Dolphin}, {Holtzman}, {Weisz}, \& {Williams}}]{McQuinn11}
{McQuinn}, K.~B.~W., {Skillman}, E.~D., {Dalcanton}, J.~J., {Dolphin}, A.~E.,
  {Holtzman}, J., {Weisz}, D.~R., \& {Williams}, B.~F. 2011, \apj, 740, 48

\bibitem[{{Meidt} {et~al.}(2012{\natexlab{a}}){Meidt}, {Schinnerer}, {Knapen},
  {Bosma}, {Athanassoula}, {Sheth}, {Buta}, {Zaritsky}, {Laurikainen},
  {Elmegreen}, {Elmegreen}, {Gadotti}, {Salo}, {Regan}, {Ho}, {Madore}, {Hinz},
  {Skibba}, {Gil de Paz}, {Mu{\~n}oz-Mateos}, {Men{\'e}ndez-Delmestre},
  {Seibert}, {Kim}, {Mizusawa}, {Laine}, \& {Comer{\'o}n}}]{Meidt12a}
{Meidt}, S.~E., {Schinnerer}, E., {Knapen}, J.~H., {Bosma}, A., {Athanassoula},
  E., {Sheth}, K., {Buta}, R.~J., {Zaritsky}, D., {Laurikainen}, E.,
  {Elmegreen}, D., {Elmegreen}, B.~G., {Gadotti}, D.~A., {Salo}, H., {Regan},
  M., {Ho}, L.~C., {Madore}, B.~F., {Hinz}, J.~L., {Skibba}, R.~A., {Gil de
  Paz}, A., {Mu{\~n}oz-Mateos}, J.-C., {Men{\'e}ndez-Delmestre}, K., {Seibert},
  M., {Kim}, T., {Mizusawa}, T., {Laine}, J., \& {Comer{\'o}n}, S.
  2012{\natexlab{a}}, \apj, 744, 17

\bibitem[{{Meidt} {et~al.}(2012{\natexlab{b}}){Meidt}, {Schinnerer},
  {Mu{\~n}oz-Mateos}, {Holwerda}, {Ho}, {Madore}, {Knapen}, {Bosma},
  {Athanassoula}, {Hinz}, {Sheth}, {Regan}, {Gil de Paz},
  {Men{\'e}ndez-Delmestre}, {Seibert}, {Kim}, {Mizusawa}, {Gadotti},
  {Laurikainen}, {Salo}, {Laine}, \& {Comer{\'o}n}}]{meidt12}
{Meidt}, S.~E., {Schinnerer}, E., {Mu{\~n}oz-Mateos}, J.-C., {Holwerda}, B.,
  {Ho}, L.~C., {Madore}, B.~F., {Knapen}, J.~H., {Bosma}, A., {Athanassoula},
  E., {Hinz}, J.~L., {Sheth}, K., {Regan}, M., {Gil de Paz}, A.,
  {Men{\'e}ndez-Delmestre}, K., {Seibert}, M., {Kim}, T., {Mizusawa}, T.,
  {Gadotti}, D.~A., {Laurikainen}, E., {Salo}, H., {Laine}, J., \&
  {Comer{\'o}n}, S. 2012{\natexlab{b}}, \apjl, 748, L30

\bibitem[{{Meixner} {et~al.}(2006){Meixner}, {Gordon}, {Indebetouw}, {Hora},
  {Whitney}, {Blum}, {Reach}, {Bernard}, {Meade}, {Babler}, {Engelbracht},
  {For}, {Misselt}, {Vijh}, {Leitherer}, {Cohen}, {Churchwell}, {Boulanger},
  {Frogel}, {Fukui}, {Gallagher}, {Gorjian}, {Harris}, {Kelly}, {Kawamura},
  {Kim}, {Latter}, {Madden}, {Markwick-Kemper}, {Mizuno}, {Mizuno}, {Mould},
  {Nota}, {Oey}, {Olsen}, {Onishi}, {Paladini}, {Panagia}, {Perez-Gonzalez},
  {Shibai}, {Sato}, {Smith}, {Staveley-Smith}, {Tielens}, {Ueta}, {van Dyk},
  {Volk}, {Werner}, \& {Zaritsky}}]{Meixner06}
{Meixner}, M., {Gordon}, K.~D., {Indebetouw}, R., {Hora}, J.~L., {Whitney}, B.,
  {Blum}, R., {Reach}, W., {Bernard}, J.-P., {Meade}, M., {Babler}, B.,
  {Engelbracht}, C.~W., {For}, B.-Q., {Misselt}, K., {Vijh}, U., {Leitherer},
  C., {Cohen}, M., {Churchwell}, E.~B., {Boulanger}, F., {Frogel}, J.~A.,
  {Fukui}, Y., {Gallagher}, J., {Gorjian}, V., {Harris}, J., {Kelly}, D.,
  {Kawamura}, A., {Kim}, S., {Latter}, W.~B., {Madden}, S., {Markwick-Kemper},
  C., {Mizuno}, A., {Mizuno}, N., {Mould}, J., {Nota}, A., {Oey}, M.~S.,
  {Olsen}, K., {Onishi}, T., {Paladini}, R., {Panagia}, N., {Perez-Gonzalez},
  P., {Shibai}, H., {Sato}, S., {Smith}, L., {Staveley-Smith}, L., {Tielens},
  A.~G.~G.~M., {Ueta}, T., {van Dyk}, S., {Volk}, K., {Werner}, M., \&
  {Zaritsky}, D. 2006, \aj, 132, 2268

\bibitem[{{Melbourne} {et~al.}(2012){Melbourne}, {Williams}, {Dalcanton},
  {Rosenfield}, {Girardi}, {Marigo}, {Weisz}, {Dolphin}, {Boyer}, {Olsen},
  {Skillman}, \& {Seth}}]{Melbourne12a}
{Melbourne}, J., {Williams}, B.~F., {Dalcanton}, J.~J., {Rosenfield}, P.,
  {Girardi}, L., {Marigo}, P., {Weisz}, D., {Dolphin}, A., {Boyer}, M.~L.,
  {Olsen}, K., {Skillman}, E., \& {Seth}, A.~C. 2012, \apj, 748, 47

\bibitem[{{Reid}(1991)}]{Reid91}
{Reid}, N. 1991, \apj, 382, 143

\bibitem[{{Rieke} {et~al.}(2004){Rieke}, {Young}, {Engelbracht}, {Kelly},
  {Low}, {Haller}, {Beeman}, {Gordon}, {Stansberry}, {Misselt}, {Cadien},
  {Morrison}, {Rivlis}, {Latter}, {Noriega-Crespo}, {Padgett}, {Stapelfeldt},
  {Hines}, {Egami}, {Muzerolle}, {Alonso-Herrero}, {Blaylock}, {Dole}, {Hinz},
  {Le Floc'h}, {Papovich}, {P{\'e}rez-Gonz{\'a}lez}, {Smith}, {Su}, {Bennett},
  {Frayer}, {Henderson}, {Lu}, {Masci}, {Pesenson}, {Rebull}, {Rho}, {Keene},
  {Stolovy}, {Wachter}, {Wheaton}, {Werner}, \& {Richards}}]{Rieke04}
{Rieke}, G.~H., {Young}, E.~T., {Engelbracht}, C.~W., {Kelly}, D.~M., {Low},
  F.~J., {Haller}, E.~E., {Beeman}, J.~W., {Gordon}, K.~D., {Stansberry},
  J.~A., {Misselt}, K.~A., {Cadien}, J., {Morrison}, J.~E., {Rivlis}, G.,
  {Latter}, W.~B., {Noriega-Crespo}, A., {Padgett}, D.~L., {Stapelfeldt},
  K.~R., {Hines}, D.~C., {Egami}, E., {Muzerolle}, J., {Alonso-Herrero}, A.,
  {Blaylock}, M., {Dole}, H., {Hinz}, J.~L., {Le Floc'h}, E., {Papovich}, C.,
  {P{\'e}rez-Gonz{\'a}lez}, P.~G., {Smith}, P.~S., {Su}, K.~Y.~L., {Bennett},
  L., {Frayer}, D.~T., {Henderson}, D., {Lu}, N., {Masci}, F., {Pesenson}, M.,
  {Rebull}, L., {Rho}, J., {Keene}, J., {Stolovy}, S., {Wachter}, S.,
  {Wheaton}, W., {Werner}, M.~W., \& {Richards}, P.~L. 2004, \apjs, 154, 25

\bibitem[{{Schirrmacher} {et~al.}(2003){Schirrmacher}, {Woitke}, \&
  {Sedlmayr}}]{Schirrmacher03}
{Schirrmacher}, V., {Woitke}, P., \& {Sedlmayr}, E. 2003, \aap, 404, 267

\bibitem[{{Sedlmayr} \& {Dominik}(1995)}]{Sedlmayr95}
{Sedlmayr}, E., \& {Dominik}, C. 1995, \ssr, 73, 211

\bibitem[{{Sheth} {et~al.}(2010){Sheth}, {Regan}, {Hinz}, {Gil de Paz},
  {Men{\'e}ndez-Delmestre}, {Mu{\~n}oz-Mateos}, {Seibert}, {Kim},
  {Laurikainen}, {Salo}, {Gadotti}, {Laine}, {Mizusawa}, {Armus},
  {Athanassoula}, {Bosma}, {Buta}, {Capak}, {Jarrett}, {Elmegreen},
  {Elmegreen}, {Knapen}, {Koda}, {Helou}, {Ho}, {Madore}, {Masters},
  {Mobasher}, {Ogle}, {Peng}, {Schinnerer}, {Surace}, {Zaritsky},
  {Comer{\'o}n}, {de Swardt}, {Meidt}, {Kasliwal}, \& {Aravena}}]{Sheth10}
{Sheth}, K., {Regan}, M., {Hinz}, J.~L., {Gil de Paz}, A.,
  {Men{\'e}ndez-Delmestre}, K., {Mu{\~n}oz-Mateos}, J.-C., {Seibert}, M.,
  {Kim}, T., {Laurikainen}, E., {Salo}, H., {Gadotti}, D.~A., {Laine}, J.,
  {Mizusawa}, T., {Armus}, L., {Athanassoula}, E., {Bosma}, A., {Buta}, R.~J.,
  {Capak}, P., {Jarrett}, T.~H., {Elmegreen}, D.~M., {Elmegreen}, B.~G.,
  {Knapen}, J.~H., {Koda}, J., {Helou}, G., {Ho}, L.~C., {Madore}, B.~F.,
  {Masters}, K.~L., {Mobasher}, B., {Ogle}, P., {Peng}, C.~Y., {Schinnerer},
  E., {Surace}, J.~A., {Zaritsky}, D., {Comer{\'o}n}, S., {de Swardt}, B.,
  {Meidt}, S.~E., {Kasliwal}, M., \& {Aravena}, M. 2010, \pasp, 122, 1397

\bibitem[{{Skrutskie} {et~al.}(2006){Skrutskie}, {Cutri}, {Stiening},
  {Weinberg}, {Schneider}, {Carpenter}, {Beichman}, {Capps}, {Chester},
  {Elias}, {Huchra}, {Liebert}, {Lonsdale}, {Monet}, {Price}, {Seitzer},
  {Jarrett}, {Kirkpatrick}, {Gizis}, {Howard}, {Evans}, {Fowler}, {Fullmer},
  {Hurt}, {Light}, {Kopan}, {Marsh}, {McCallon}, {Tam}, {Van Dyk}, \&
  {Wheelock}}]{Skrutskie06}
{Skrutskie}, M.~F., {Cutri}, R.~M., {Stiening}, R., {Weinberg}, M.~D.,
  {Schneider}, S., {Carpenter}, J.~M., {Beichman}, C., {Capps}, R., {Chester},
  T., {Elias}, J., {Huchra}, J., {Liebert}, J., {Lonsdale}, C., {Monet}, D.~G.,
  {Price}, S., {Seitzer}, P., {Jarrett}, T., {Kirkpatrick}, J.~D., {Gizis},
  J.~E., {Howard}, E., {Evans}, T., {Fowler}, J., {Fullmer}, L., {Hurt}, R.,
  {Light}, R., {Kopan}, E.~L., {Marsh}, K.~A., {McCallon}, H.~L., {Tam}, R.,
  {Van Dyk}, S., \& {Wheelock}, S. 2006, \aj, 131, 1163

\bibitem[{{van Loon} {et~al.}(2008){van Loon}, {Boyer}, \&
  {McDonald}}]{van-Loon08}
{van Loon}, J.~T., {Boyer}, M.~L., \& {McDonald}, I. 2008, \apjl, 680, L49

\bibitem[{{van Loon} {et~al.}(2005){van Loon}, {Marshall}, \&
  {Zijlstra}}]{van-Loon05}
{van Loon}, J.~{\relax Th}., {Marshall}, J.~R., \& {Zijlstra}, A.~A. 2005,
  \aap, 442, 597

\bibitem[{{Verley} {et~al.}(2009){Verley}, {Corbelli}, {Giovanardi}, \&
  {Hunt}}]{Verley09}
{Verley}, S., {Corbelli}, E., {Giovanardi}, C., \& {Hunt}, L.~K. 2009, \aap,
  493, 453

\bibitem[{{Winters} {et~al.}(2000){Winters}, {Le Bertre}, {Jeong}, {Helling},
  \& {Sedlmayr}}]{Winters00}
{Winters}, J.~M., {Le Bertre}, T., {Jeong}, K.~S., {Helling}, C., \&
  {Sedlmayr}, E. 2000, \aap, 361, 641

\bibitem[{{Winters} {et~al.}(2003){Winters}, {Le Bertre}, {Jeong}, {Nyman}, \&
  {Epchtein}}]{Winters03}
{Winters}, J.~M., {Le Bertre}, T., {Jeong}, K.~S., {Nyman}, L.-{\AA}., \&
  {Epchtein}, N. 2003, \aap, 409, 715

\bibitem[{{Woitke}(2006)}]{Woitke06}
{Woitke}, P. 2006, \aap, 452, 537

\bibitem[{{Woitke}(2007)}]{Woitke07}
{Woitke}, P. 2007, in Astronomical Society of the Pacific Conference Series,
  Vol. 378, Why Galaxies Care About AGB Stars: Their Importance as Actors and
  Probes, ed. F.~{Kerschbaum}, C.~{Charbonnel}, \& R.~F. {Wing}, 156

\bibitem[{{Zaritsky} {et~al.}(2002){Zaritsky}, {Harris}, {Thompson}, {Grebel},
  \& {Massey}}]{Zaritsky02}
{Zaritsky}, D., {Harris}, J., {Thompson}, I.~B., {Grebel}, E.~K., \& {Massey},
  P. 2002, \aj, 123, 855

\bibitem[{{Zibetti} {et~al.}(2009){Zibetti}, {Charlot}, \& {Rix}}]{Zibetti09}
{Zibetti}, S., {Charlot}, S., \& {Rix}, H.-W. 2009, \mnras, 400, 1181

\end{thebibliography}

\clearpage

%\begin{landscape}

\center
\begin{deluxetable*}{lcccccccccc}
\tabletypesize{\scriptsize}
\tablecaption{Integrated Flux Density of the SMC \& LMC from $1 -24$~\um\ \\and Percentage of that Flux Contributed by TP-AGB and RSG Stars\label{tab:results}}
\tablehead{\colhead{Type} & \colhead{Stars\tablenotemark{a}}  & \colhead{Fgnd\tablenotemark{b}} &\colhead{1.2\um \tablenotemark{c}} & \colhead{2.2\um\tablenotemark{c}}
& \colhead{3.5\um\tablenotemark{d}} & \colhead{3.6\um\tablenotemark{e}} & \colhead{4.5\um\tablenotemark{e}}
 & \colhead{5.8\um\tablenotemark{e}} & \colhead{8\um\tablenotemark{e}} & \colhead{24\um\tablenotemark{e}}}
\startdata
\hline
SMC	&&&&&&&&\\
&&&\multicolumn{8}{c}{Integrated Flux [Jy]} \\
\cline{4-11}\\
&&& $670\pm 80$\tablenotemark{f} & $525\pm 65$\tablenotemark{f} & $280 \pm 40$\tablenotemark{f} & $300 \pm 20$\tablenotemark{g} & $220 \pm 10$\tablenotemark{g} & $220 \pm 10$\tablenotemark{g} & $200 \pm 10$\tablenotemark{g} & $350 \pm 10$\tablenotemark{g}		\\
&&&&&&&&\\
&&&\multicolumn{8}{c}{Fractional Contribution [\%]} \\
\cline{4-11}\\
O-AGB&3353  & 4 & $5.7_{-0.7}^{+1.1}$ & $8.4_{-1.0}^{+1.5}$ &
$8.0_{-1.1}^{+1.5}$ & $7.5_{-0.5}^{+1.1}$ & $6.0_{-0.3}^{+0.8}$ &
$4.5_{-0.2}^{+0.6}$ & $3.1_{-0.2}^{+0.4}$ & $0.14_{-0.01}^{+0.01}$  \\
C-AGB&1597 & $<1$ & $3.0_{-0.4}^{+0.4}$ & $6.6_{-0.8}^{+0.9}$ 
& $9.1_{-1.3}^{+1.4}$ & $8.5_{-0.6}^{+0.8}$ & $7.1_{-0.3}^{+0.6}$ 
& $5.3_{-0.2}^{+0.4}$ & $4.8_{-0.2}^{+0.4}$ & $0.31_{-0.01}^{+0.02}$ \\
x-AGB&309  & $<1$ & $0.2_{-0.1}^{+0.1}$ & $1.2_{-0.1}^{+0.1}$
& $4.8_{-0.7}^{+0.7}$ & $4.5_{-0.3}^{+0.3}$ & $6.5_{-0.3}^{+0.3}$ 
& $6.9_{-0.3}^{+0.3}$ & $7.6_{-0.4}^{+0.4}$ & $1.30_{-0.04}^{+0.04}$ \\
RSG &2340  &598   & $8.9_{-1.1}^{+1.1}$ & $10.9_{-1.4}^{+1.4}$ 
& $10.0_{-1.4}^{+1.4}$ & $9.4_{-0.6}^{+0.6}$ & $7.6_{-0.3}^{+0.3}$ 
& $5.5_{-0.3}^{+0.3}$  & $3.7_{-0.2}^{+0.2}$ & $0.33_{-0.01}^{+0.01}$ \\
\hline
Total& 7599 & 602  & $17.8_{-1.3}^{+1.8}$ & $27.1_{-1.9}^{+2.7}$ 
& $31.9_{-2.3}^{+3.4}$ & $29.8_{-1.0}^{+2.5}$ & $27.2_{-0.6}^{+2.3}$ 
& $22.3_{-0.5}^{+2.0}$ & $19.3_{-0.5}^{+2.1}$ & $2.1_{-0.1}^{+0.2}$ \\ 
&&&&&&&&\\                                                                                                                                                
\hline\hline
LMC&&&&&&&&\\
&&&\multicolumn{8}{c}{Integrated Flux [Jy]} \\
\cline{4-11}\\
 &&& $4520 \pm 650$\tablenotemark{f} & $3770 \pm 540$\tablenotemark{f} & $2190 \pm 300$\tablenotemark{f}  & $ 2080 \pm 100$\tablenotemark{h} & $1350 \pm 70 $\tablenotemark{h} & $1850 \pm 200 $\tablenotemark{h} & $5980 \pm 300 $\tablenotemark{h} & $7640 \pm 200$\tablenotemark{h} \\ 
&&&&&&&&\\
&&&\multicolumn{8}{c}{Fractional Contribution [\%]} \\
\cline{4-11}\\                                                 
O-AGB&15635 & $<1$ & $6.8_{-1.0}^{+2.6}$ & $9.9_{-1.1}^{+3.1}$ 
& $8.8_{-0.8}^{+3.1}$  & $9.3_{-0.5}^{+3.2}$ & $8.3_{-0.4}^{+3.1}$ 
& $4.6_{-0.5}^{+1.7}$  & $0.9_{-0.1}^{+0.3}$ & $0.11_{-0.01}^{+0.22}$ \\
C-AGB&5628  & $<1$ & $2.6_{-0.4}^{+0.4}$ & $6.0_{-0.6}^{+0.7}$ 
& $7.6_{-0.7}^{+0.7}$    & $8.1_{-0.4}^{+0.5}$ & $7.3_{-0.4}^{+0.4}$ 
& $3.7_{-0.4}^{+0.4}$    & $1.0_{-0.1}^{+0.1}$ &
$0.11_{-0.01}^{+0.01}$  \\
x-AGB&989   & $<1$ & $0.1_{-0.1}^{+0.1}$ & $0.7_{-0.1}^{+0.1}$ 
& $2.9_{-0.3}^{+0.3}$  & $3.0_{-0.1}^{+0.2}$ & $5.2_{-0.3}^{+0.3}$ 
& $4.3_{-0.5}^{+0.5}$  & $1.4_{-0.1}^{+0.1}$ & $0.35_{-0.01}^{+0.01}$ \\
RSG&4361 & 13  & $5.5_{-0.8}^{+0.8}$ & $6.9_{-0.7}^{+0.7}$ 
& $6.0_{-0.5}^{+0.5}$  & $6.3_{-0.3}^{+0.3}$ & $5.5_{-0.3}^{+0.3}$ 
& $3.0_{-0.3}^{+0.3}$  & $0.6_{-0.1}^{+0.1}$ & $0.16_{-0.01}^{+0.01}$ \\
\hline
Total&26613 & 13 & $15.1_{-1.3}^{+2.8}$ & $23.5_{-1.4}^{+3.3}$ 
& $25.2_{-1.2}^{+3.3}$ & $26.6_{-0.7}^{+3.3}$ & $26.3_{-0.7}^{+3.2}$ 
& $15.6_{-0.9}^{+1.9}$ & $3.9_{-0.1}^{+0.4}$  & $0.7_{-0.1}^{+0.2}$ \\
\enddata
\tablenotetext{a}{Number of stars of given type detected at 3.6 \um; the number varies slightly with wavelength}
\tablenotetext{b}{Number of Milky Way foreground stars as estimated by TRILEGAL \citep{Girardi05}}
\tablenotetext{c}{Integrated Flux from COBE/DIRBE (at 1.25 and 2.2 \um), Stellar Flux from 2MASS (at 1.2 and 2.2 \um)}
\tablenotetext{d}{Integrated Flux from COBE/DIRBE (at 3.5 \um),
  Stellar Flux from {\it Spitzer} (at 3.6 \um)}
\tablenotetext{e}{Integrated Flux from {\it Spitzer}, Stellar Flux
  from {\it Spitzer}}
\tablenotetext{f}{From \citet{Israel10}}
\tablenotetext{g}{From \citet{Gordon11}}
\tablenotetext{h}{From Gordon et al. (in preparation)}
\end{deluxetable*}

%\end{landscape}

\clearpage

\end{document}